\RequirePackage{fix-cm}

\documentclass[smallextended]{svjour3}       

\usepackage{amsfonts,amssymb,amsmath}
\usepackage{graphics,graphicx,epsfig}
\usepackage[dvipsnames]{xcolor}
\usepackage[normalem]{ulem}
\usepackage{comment}
\usepackage{epstopdf}
\usepackage{graphicx}
\usepackage{subfigure}
\usepackage{marvosym}
\usepackage{hyperref}
\usepackage{geometry}
\usepackage{cite}
\usepackage [latin1]{inputenc}

\def\be{\begin{eqnarray}}
\def\ee{\end{eqnarray}}
\def\ba{\begin{array}{l}}
\def\ea{\end{array}}

\begin{document}
\title{Detection of genuine tripartite entanglement based on Bloch representation of density matrices}
\author{Hui Zhao$^1$ \and Yu-Qiu Liu $^1$  \and Naihuan Jing$^{2}$  \and Zhi-Xi Wang$^3$ \and Shao-Ming Fei$^{3,4}$}
\institute{ \Letter~~corresponding author:Hui Zhao \at
                     zhaohui@bjut.edu.cn \and
              \Letter~~Yu-Qiu Liu \at
                     1091960127@qq.com \and
            \Letter~~    Naihuan Jing\at
                    ~~~ jing@math.ncsu.edu \and
              \Letter~~     Zhi-Xi Wang\at
                     ~~~wangzhx@cnu.edu.cn \and
              \Letter~~     Shao-Ming Fei\at
                     ~~~feishm@cnu.edu.cn
           \and
\at 1 Department of Mathematics, Faculty of Science, Beijing University of Technology, Beijing 100124, China\\
\at 2 Department of Mathematics, North Carolina State University, Raleigh, NC 27695, USA\\
\at 3 School of Mathematical Sciences,  Capital Normal University,  Beijing 100048,  China\\
\at 4 Max-Planck-Institute for Mathematics in the Sciences, Leipzig 04103, Germany
}

\maketitle
\begin{abstract}
\baselineskip18pt
We study the genuine multipartite entanglement in tripartite quantum systems.
By using the Schmidt decomposition and local unitary transformation, we convert the general states to simpler forms and consider
 certain matrices from correlation tensors in the Bloch representation of the simplified density matrices.
Using these special matrices we obtain new criteria for genuine multipartite entanglement. Detail examples show that our criteria
are able to detect more tripartite entangled and genuine tripartite entangled states than some existing criteria.

\end{abstract}

\section{Introduction}
Quantum entanglement is a key resource in varieties of information processing such as quantum cryptography \cite{ea}, teleportation \cite{bbj} and dense coding \cite{bw} etc. Among various types of entanglement, genuine multipartite entanglement stands out to be advantageous in many applications \cite{hr}. Therefore, measuring and detection of genuine multipartite entanglement are important tasks.

Denote by $H_{i}^{d_{i}}$ the $i$th $d_{i}$-dimensional Hilbert space.
A quantum pure tripartite state $|\varphi\rangle\in H_{1}^{d_{1}}\otimes H_{2}^{d_{2}}\otimes H_{3}^{d_{3}}$ is called separable under the bipartition $f|gh$ if $|\varphi\rangle$ can be written as $|\varphi\rangle=|\varphi_{f}\rangle\otimes|\varphi_{gh}\rangle$, where $|\varphi_{f}\rangle$ and $|\varphi_{gh}\rangle$ are the respective
 pure states in $H_{f}^{d_{f}}$ and $H_{g}^{d_{g}}\otimes H_{h}^{d_{h}}(f\neq g\neq h\in\{1,2,3\})$.
Biseparability can then be defined similarly for multipartite systems.  A pure non-biseparable state is called genuine multipartite entangled (GME). For any mixed state $\rho=\sum p_{i}\rho_{i}$, where $0<p_{i}\leq1$ and $\sum p_{i}=1$, if $\rho$ can not be written as a pure state decomposition of biseparable ones, then $\rho$ is called GME.

In recent years, quite a few methods have been proposed to detect genuine entanglement. Many criteria detecting GME were based on GME-concurrence \cite{mcc,cmc,ygy}, entanglement witness \cite{bgl,wkb} and realignment of density matrices \cite{cw,lws}. The authors in \cite{hs} introduced a general framework that allows to construct witnesses for genuine multipartite entanglement directly from positive maps. By using the norms of correlation vector, the authors in \cite{vh} presented an approaches to detect both GME and not-fully separable in multipartite arbitrary dimensional quantum systems. In \cite{mlp}, a criterion for GME was derived by considering the sum of squares of all possible bipartite correlations and using the monogamy relations of quantum correlations. The sufficient criterion for GME in tripartite and multipartite system have also been presented in \cite{ljw,zzn}.

In this paper, we first transform a general state into a simpler form by the Schmidt decomposition and local unitary (LU) transformation. Then we construct some special matrices to extract information of GME. In terms of
the trace norm of linear combination of these matrices, we derive new criteria on GME.
Let us briefly review the Schmidt decomposition  \cite{man} and LU equivalence  \cite{JYZ,CCZ} to explain our approach.
Suppose $|\varphi\rangle$ is a pure state of a composite system $H_1^{d_1}\otimes H_2^{d_2}$. Then there exist orthonormal states $|i_a\rangle$ for system $H_1^{d_1}$ and orthonormal states $|i_b\rangle$ for system $H_2^{d_2}$ such that $|\varphi\rangle$ has the Schmidt decomposition $|\varphi\rangle=\sum_i\tau_i|i_a\rangle|i_b\rangle$, where $\tau_i$ are non-negative real numbers satisfying $\sum_i \tau_i^2=1$, known as the
Schmidt coefficients \cite{man}. For a bipartite state on $H_1^{d_1}\otimes H_2^{d_2}$, $\rho$ can be expressed as $\rho=\frac{1}{d_1d_2}I_{d_1}\otimes I_{d_2}+\sum\limits_{i}t_{i}^{1}\lambda_{i}^{(1)}\otimes I_{d_2}+\sum\limits_{j}t_{j}^{2}I_{d_1}\otimes\lambda_{j}^{(2)}
+\sum\limits_{ij}t_{ij}\lambda_{i}^{(1)}\otimes \lambda_{j}^{(2)}$, where $\{\lambda_i^{(k)}, 0\leq i\leq d_k^2-1, k=1,2\}$ are the generators of the special unitary Lie group $SU(d_k)$.
Define the matrix $N=[t_{ij}]_{(d_1^2-1)\times (d_2^2-1)}$.
Let $\rho'=(U_1\otimes U_2)\rho(U_1^{\dagger}\otimes U_2^{\dagger})$ be another bipartite state on $H_1^{d_1}\otimes H_2^{d_2}$, where $U_1$ and $U_2$ are unitary matrices. One sees that $\sum\limits_{ij}t_{ij}(\lambda_{i}^{(1)})^{U_1}\otimes (\lambda_{j}^{(2)})^{U_2}=\sum\limits_{ij}(\sum\limits_{i'j'}a_{i'i}t_{i'j'}b_{j'j})
\lambda_{i}^{(1)}\otimes \lambda_{j}^{(2)}$ with two real orthogonal matrices $A=[a_{ij}]\in O(d_1^2-1)$ and $B=[b_{ij}]\in O(d_2^2-1)$. Therefore, $N(\rho')=A^tN(\rho)B$ \cite{JYZ}. Let $\|\cdot\|_{tr}$ stand for the trace norm defined by $\|N\|_{tr}=\sum_{i}\sigma_{i}=\mbox Tr\sqrt{N^{\dagger}N}$, $N\in\mathbb{R}^{m\times n}$, where $\sigma_{i}$ $(i=1,2,\cdot\cdot\cdot, \mbox min\{m,n\})$ are the singular values of the matrix $N$. Then we have $\|N(\rho')\|_{tr}=\|N(\rho)\|_{tr}$ due to the fact that the singular values of a rectangular matrix
$N$ are the same as those of $B^tNA$ when $A, B$ are orthogonal. Thus the study of $\rho$ can be translated into that of $\rho'$.
Using this idea, we are able to construct some useful invariants of GME.

The paper is organized as follows: in Section 2, we present new separability criteria to detect GME for $(2\times2\times 2)$-dimensional quantum states, a detailed example shows that our theorem is more effective than previous available results. In Section 3, we generalize these criteria to $(d\times d\times d)$-dimensional systems. Comments and conclusions are given in Section 4.

\section{GME for $(2\times2\times 2)$-dimensional quantum states}
We first consider the separability and GME of three qubit states. Denote by $\lambda_{i}^{(j)}$ $(i=1, 2, 3)$ the standard Pauli spin matrices $\sigma_3$, $\sigma_1$ and $\sigma_2$, associated with
the $j$th qubit, respectively.
A general three qubit state $\rho$ can be written in the following Bloch representation,
\begin{equation}\label{1}
\begin{split}
\rho=&\frac{1}{8}(I\otimes I\otimes I+\sum\limits_{i=1}^{3}t_{i}^{1}\lambda_{i}^{(1)}\otimes I\otimes I+\sum\limits_{j=1}^{3}t_{j}^{2}I\otimes \lambda_{j}^{(2)}\otimes I+\sum\limits_{k=1}^{3}t_{k}^{3}I\otimes I\otimes \lambda_{k}^{(3)}\\
&+\sum\limits_{i,j=1}^{3}t_{ij}^{12}\lambda_{i}^{(1)}\otimes \lambda_{j}^{(2)}\otimes I+\sum\limits_{i,k=1}^{3}t_{ik}^{13}\lambda_{i}^{(1)}\otimes I\otimes \lambda_{k}^{(3)}+\sum\limits_{j,k=1}^{3}t_{jk}^{23}I\otimes \lambda_{j}^{(2)}\otimes \lambda_{k}^{(3)}\\
&+\sum\limits_{i,j,k=1}^{3}t_{ijk}\lambda_{i}^{(1)}\otimes \lambda_{j}^{(2)}\otimes \lambda_{k}^{(3)}),
\end{split}
\end{equation}
where $t_{i}^{1}=Tr(\rho\lambda_{i}^{(1)}\otimes I\otimes I)$, $t_{j}^{2}=Tr(\rho I\otimes \lambda_{j}^{(2)}\otimes I)$, $t_{k}^{3}=Tr(\rho I\otimes I\otimes \lambda_{k}^{(3)})$, $t_{ij}^{12}=\mbox Tr(\rho\lambda_{i}^{(1)}\otimes\lambda_{j}^{(2)}\otimes I)$, $t_{ik}^{13}=\mbox Tr(\rho\lambda_{i}^{(1)}\otimes I\otimes \lambda_{k}^{(3)})$, $t_{jk}^{23}=\mbox Tr(\rho I\otimes\lambda_{j}^{(2)}\otimes\lambda_{k}^{(3)})$ and $t_{ijk}=\mbox Tr(\rho\lambda_{i}^{(1)}\otimes\lambda_{j}^{(2)}\otimes\lambda_{k}^{(3)})$.
Let $T^{1|23}_{1}, T^{1|23}_{2}, T^{1|23}_{3}, T^{2|13}, T^{3|12}_{1}, T^{3|12}_{2}$ and $T^{3|12}_{3}$ denote the matrices with entries $t_{1jk}, t_{2jk}, t_{3jk}, t_{i1k}, t_{ij1}, t_{ij2}$ and $t_{ij3}$ $(i,j,k=1,2,3)$, respectively.
For example,
\begin{equation}\label{2}
T^{1|23}_{1}=\left[
                        \begin{array}{ccc}
                          t_{111} & t_{121} & t_{131} \\
                          t_{112} & t_{122} & t_{132} \\
                          t_{113} & t_{123} & t_{133} \\
                        \end{array}
                      \right],~~~
T^{2|13}=\left[
                \begin{array}{ccc}
                  t_{111} & t_{211} & t_{311} \\
                  t_{112} & t_{212} & t_{312} \\
                  t_{113} & t_{213} & t_{313} \\
                \end{array}
              \right].
\end{equation}
Set $N^{1|23}=15T_{1}^{1|23}+T_{2}^{1|23}+T_{3}^{1|23}$, $N^{2|13}=4T^{2|13}$, $N^{3|12}=15T_{1}^{3|12}+T_{2}^{3|12}+T_{3}^{3|12}$ and $T(\rho)=\frac{1}{3}(\|N^{1|23}\|_{tr}+\|N^{2|13}\|_{tr}+\|N^{3|12}\|_{tr})$.

{\bf Remark 1.} There are two kinds of genuine three-qubit entangled pure states under stochastic local operations and communication (SLOCC), namely, the GHZ state and W state. Mixing the GHZ or W states with white noise, one can obtain different upper bounds of $\|N^{f|gh}\|_{tr}$ to detect their entanglement. We find that $T(\rho)=\frac{1}{3}(\|N^{1|23}\|_{tr}+\|N^{2|13}\|_{tr}+\|N^{3|12}\|_{tr})$ can detect more genuine tripartite entangled states in $d\times d\times d$ dimensional systems.

Note that $\|N^{1|23}\|_{tr}$ is invariant under local unitary transformations. Suppose $\rho'=(I\otimes U_2\otimes U_3)\rho(I\otimes U_2^{\dagger}\otimes U_3^{\dagger})$, where $U_2, U_3\in U(2)$, $U_2\lambda_i^{(2)}U_2^{\dagger}=\sum_{j=1}^3a_{ij}\lambda_j^{(2)}$ and $U_3\lambda_i^{(3)}U_3^{\dagger}=\sum_{j=1}^3b_{ij}\lambda_j^{(3)}$ for some coefficients $a_{ij}$ and $b_{ij}$. By \cite[Lemma 2.1]{JYZ} one has that $A=(a_{ij}), B=(b_{ij})\in O(3)$ and
\begin{equation}\label{3}
T^{1|23}_1(\rho')=B^tT^{1|23}_1(\rho)A, \qquad T^{1|23}_2(\rho')=B^tT^{1|23}_2(\rho)A, \qquad T^{1|23}_3(\rho')=B^tT^{1|23}_3(\rho)A.
\end{equation}
Then we have
\begin{equation}\label{4}
N^{1|23}(\rho')=15B^tT^{1|23}_1(\rho)A+B^tT^{1|23}_2(\rho)A+B^tT^{1|23}_3(\rho)A=B^tN^{1|23}(\rho)A.
\end{equation}
The singular value decomposition of $N^{1|23}(\rho)$ is $N^{1|23}(\rho)=UDV$, where $U$ and $V$ are unitary matrices, $D$ is a diagonal matrix with singular value of $N^{1|23}(\rho)$. Then $N^{1|23}(\rho')=B^tUDVA$ shares the same matrix $D$ with $N^{1|23}(\rho)$. Thus the singular values of $N^{1|23}(\rho)$ and $N^{1|23}(\rho')$ are the same. Therefore, $\|N^{1|23}(\rho')\|_{tr}=\|N^{1|23}(\rho)\|_{tr}$.
Due to invariance of the trace norm under local unitary transformations, we can simplify form of the density matrix to study the entanglement of the tripartite quantum states.

We first consider biseparable pure states. If $\rho=|\varphi\rangle\langle\varphi|$ is $1|23$ separable under the bipartition of the first qubit and the last two qubits. i.e.,
$|\varphi_{1|23}\rangle=|\varphi_{1}\rangle\otimes|\varphi_{23}\rangle\in H_{1}^{2}\otimes H_{23}^{4}$.
From Schmidt decomposition, we have
\begin{equation}\label{5}
|\varphi_{1|23}\rangle=\tau_{0}|0a\rangle+\tau_{1}|1b\rangle,
\end{equation}
with $|\tau_{0}|^{2}+|\tau_{1}|^{2}=1$.
Taking into account local unitary equivalence in $H_2^2\otimes H_3^2$, when $|\varphi_{23}\rangle\in H_{23}^{4}$ is separable, we can transform $\{|a\rangle, |b\rangle\}$ into two orthonormal basis elements which constitute separable states : (i) $\{|a\rangle, |b\rangle\}=\{|00\rangle, |01\rangle\}$, i.e. $|\varphi_{1|23}\rangle=|\varphi_{1}\rangle\otimes|\varphi_{2}\rangle\otimes|\varphi_{3}\rangle$ is fully separable; when $|\varphi_{23}\rangle\in H_{23}^{4}$ is entangled, we can transform $\{|a\rangle, |b\rangle\}$ into two orthonormal basis elements which constitute entangled states :
(ii) $\{|a\rangle, |b\rangle\}=\{|00\rangle, |11\rangle\}$.
The matrices $T_{1}^{1|23}$, $T_{2}^{1|23}$  and $T_{3}^{1|23}$ are given by, respectively,
\begin{align}\label{6}
&(i):T_{1}^{1|23}=\left[
                                \begin{array}{ccc}
                                  \tau_{0}^{2}+\tau_{1}^{2} & 0 & 0 \\
                                   0 & 0 & 0 \\
                                  0 & 0 & 0 \\
                                \end{array}
                              \right],~~~
T_{2}^{1|23}=\left[
               \begin{array}{ccc}
                 0 & 0 & 0\\
                 2\tau_{0}\tau_{1} & 0 & 0 \\
                 0 & 0 & 0 \\
               \end{array}
             \right],~~~
T_{3}^{1|23}=\left[
               \begin{array}{ccc}
                 0 & 0 & 0\\
                 0 & 0 & 0 \\
                 -2\tau_{0}\tau_{1} & 0 & 0 \\
               \end{array}
             \right],\\
&(ii):T_{1}^{1|23}=\left[
                                \begin{array}{ccc}
                                  \tau_{0}^{2}-\tau_{1}^{2} & 0 & 0 \\
                                  0 & 0& 0 \\
                                  0 & 0 & 0
                                \end{array}
                              \right],~~
T_{2}^{1|23}=\left[
               \begin{array}{ccc}
                 0 & 0 & 0\\
                 0 & 2\tau_{0}\tau_{1} & 0 \\
                 0 & 0 & -2\tau_{0}\tau_{1}
               \end{array}
             \right],~~
T_{3}^{1|23}=\left[
               \begin{array}{ccc}
                 0 & 0 & 0\\
                 0 & 0 & -2\tau_{0}\tau_{1} \\
                 0 & -2\tau_{0}\tau_{1} & 0
               \end{array}
             \right].
\end{align}

We find that after Schmidt
decomposition and LU equivalence, the matrices (2) constructed by correlation tensors have been greatly simplified to (5) and (6), then we have the following separability criterion.

\begin{lemma}\label{lemma:1}If the state $\rho\in H_{1}^{2}\otimes H_{2}^{2}\otimes H_{3}^{2}$ is a bipartite separable pure state, corresponding to the case (i) and (ii), we have\\
(1) If $\rho$ is separable under bipartition $1|23$, then $\|N^{1|23}\|_{tr}\leq\sqrt{227}$~~or~~$\sqrt{233}$;\\
(2) If $\rho$ is separable under bipartition $2|13$, then $\|N^{2|13}\|_{tr}\leq12$~~or~~$4$;\\
(3) If $\rho$ is separable under bipartition $3|12$, then $\|N^{3|12}\|_{tr}\leq\sqrt{227}$~~or~~$\sqrt{233}$;
\end{lemma}
{\it Proof}~~(1) If a pure tripartite qubit state $|\varphi\rangle\in H_{1}^{2}\otimes H_{2}^{2}\otimes H_{3}^{2}$ is separable under the bipartition $1|23$, then for the case (i),
\begin{equation}\label{8}
\|N^{1|23}\|_{tr}=\sqrt{15^2(\tau_{0}^{2}+\tau_{1}^{2})^{2}+8\tau_{0}^{2}\tau_{1}^{2}}\leq\sqrt{227}.
\end{equation}
And for the case (ii),
\begin{equation}\label{9}
\begin{split}
\|N^{1|23}\|_{tr}&=15\sqrt{(\tau_{0}^{2}-\tau_{1}^{2})^{2}}+4\sqrt{2}\tau_{0}\tau_{1}\\
&=15\sqrt{(1-2\tau_{1}^{2})^{2}}+4\sqrt{2}\tau_{1}\sqrt{1-\tau_{1}^{2}}\\
&\leq\sqrt{233},
\end{split}
\end{equation}
where the upper bound is obtained by taking the extreme value of the function with independent variable $\tau_{1}$.\\
(2) If $\rho$ is separable under bipartition $2|13$, for the first case,
\begin{equation}\label{10}
T^{2|13}=\left[
                  \begin{array}{ccc}
                  \tau_{0}^{2}+\tau_{1}^{2} & 0 & 0 \\
                   0 & 2\tau_{0}\tau_{1} & 0 \\
                   0 & 0 & -2\tau_{0}\tau_{1} \\
                   \end{array}
                   \right],
\end{equation}
we have
\begin{equation}\label{11}
\|N^{2|13}\|_{tr}=4(\tau_{0}^{2}+\tau_{1}^{2}+4\tau_{0}\tau_{1})\leq12.
\end{equation}
And for the second case,
\begin{align}\label{12}
T^{2|13}=\left[
                   \begin{array}{ccc}
                   \tau_{0}^{2}-\tau_{1}^{2} & 0 & 0 \\
                   0 & 0& 0 \\
                   0 & 0 & 0 \\
                   \end{array}
                   \right],
\end{align}
then $$\|N^{2|13}\|_{tr}=4\sqrt{(\tau_{0}^{2}-\tau_{1}^{2})^{2}}\leq4.$$
(3) Using similar method, if $\rho$ is separable under bipartition $3|12$, $\|N^{3|12}\|_{tr}\leq\sqrt{227}$ and $\sqrt{233}$ with respect to the case (i) and (ii), respectively.
\qed

{\bf Remark 2.} Under Schmidt decomposition a general state can be written as $|\varphi_{1|23}\rangle=|\varphi_{1}\rangle\otimes|\varphi_{23}\rangle=\tau_{0}|0a\rangle+\tau_{1}|1b\rangle$. Under LU equivalence we can choose $\{|a\rangle, |b\rangle\}=\{|00\rangle, |01\rangle\}$ in the case (i) to transform $\{|a\rangle, |b\rangle\}$ into two orthonormal bases which constitute
separable states. Then $|\varphi_{23}\rangle$ must be a separable state such that $|\varphi_{23}\rangle=|\varphi_2\rangle\otimes|\varphi_{3}\rangle$, which implies that $|\varphi\rangle$ is fully separable. Hence, we have $\|N^{1|23}\|_{tr}\leq\sqrt{227}$. Consequently, if $\|N^{1|23}\|_{tr}>\sqrt{227}$, $\rho$ is not fully separable. Therefore, Lemma 1 can also be used to detect the fully separability.

\begin{theorem}\label{thm:1} For a tripartite qubit mixed state $\rho$, if
\begin{equation}\label{12}
T(\rho)=\frac{1}{3}(\|N^{1|23}\|_{tr}+\|N^{2|13}\|_{tr}+
\|N^{3|12}\|_{tr})>\sqrt{233},
\end{equation}
then $\rho$ is genuine multipartite entangled.
\end{theorem}
{\it Proof}~~For a mixed state $\rho=\sum p_{i}\rho_{i},\sum p_{i}=1$, if $\rho$ is biseparable, by using Lemma 1, we have
\begin{equation*}
\begin{split}
T(\rho)=&\frac{1}{3}(\|N^{1|23}(\rho)\|_{tr}+\|N^{2|13}(\rho)\|_{tr}+\|N^{3|12}(\rho)\|_{tr})\\
\leq&\frac{1}{3}\sum p_{i}(\|N^{1|23}(\rho_{i})\|_{tr}+\|N^{2|13}(\rho_{i})\|_{tr}+\|N^{3|12}(\rho_{i})\|_{tr})\\
\leq&\frac{1}{3}(\sqrt{233}+\sqrt{233}+\sqrt{233})\\
=&\sqrt{233}.
\end{split}
\end{equation*}
Consequently, if $T(\rho)>\sqrt{233}$, $\rho$ is GME.
\qed

\textit{\textbf{Example 1}} Consider the mixture of the W state with maximally mixed state,
\begin{equation}\label{13}
\rho=\frac{1-x}{8}I_{8}+x|W\rangle\langle W|,
\end{equation}
where $|W\rangle=\frac{1}{\sqrt{3}}(|001\rangle+|010\rangle+|100\rangle)$, $x\in[0,1]$, $I_{8}$ is $8\times 8$ identity matrix. By calculation, we have that $\|N^{1|23}\|_{tr}=(\sqrt{\frac{2941}{18}-\frac{5\sqrt{5657}}{6}}+\sqrt{\frac{2941}{18}+\frac{5\sqrt{5657}}{6}}+10)x$.
Using Lemma 1, we have when $\|N^{1|23}\|_{tr}>\sqrt{227}$, $\rho$ is not fully separable; when $\|N^{1|23}\|_{tr}>\sqrt{233}$, $\rho$ is not separable under bipartition $1|23$. Therefore, we have $\rho$ is not fully separable for $0.4296<x\leq1$ and not separable under bipartition $1|23$ for $0.4352<x\leq1$.
In \cite{chl}, $\rho$ was detected as entangled for $0.619<x\leq1$. This shows that Lemma 1 detects more entangled states.

By using Theorem 1, we have $f_{1}(x)=T(\rho)-\sqrt{233}=\frac{1}{3}[2(\sqrt{\frac{2941}{18}-\frac{5\sqrt{5657}}{6}}+\sqrt{\frac{2941}{18}+\frac{5\sqrt{5657}}{6}}+10)+\frac{28}{3}]x-\sqrt{233}$,
$\rho$ is a GME state if $f_{1}(x)>0$, namely, $0.5762<x\leq1$. Set $f_2(x)=(\sqrt{\frac{17}{9}}+2\sqrt{\frac{8}{9}})x-\frac{6+\sqrt{3}}{3}$, using Theorem 2 in \cite{vh}, $f_{2}(x)>0$ is used to detect GME for $0.791<x\leq1$. Set $f_3(x)=\frac{1}{12}(\sqrt{66}x-6)$, using Theorem 2 in \cite{ljw}, $\rho$ is GME if $f_{3}(x)>0$, i.e., $0.7385<x\leq1$. The comparison is shown in Fig.1, where our result is able to detect more GME states.
\begin{figure}
  \centering
  \includegraphics[width=8cm]{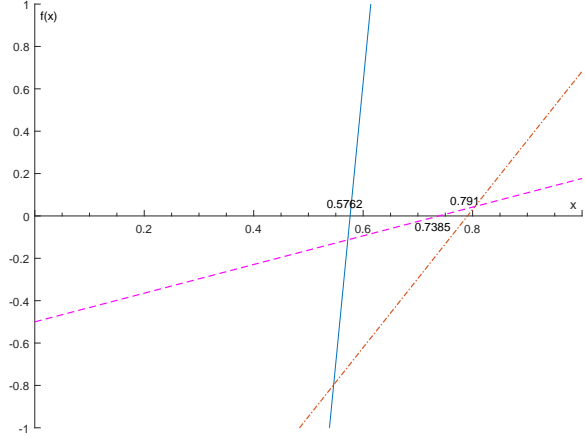}\\
  \caption{Detect GME for $\rho$ in Example 1, $f_1(x)$ from our Theorem 1 (solid straight line), $f_2(x)$ from Theorem 2 in \cite{vh}(dash-dot straight line) and $f_3(x)$ from Theorem 2 in \cite{ljw}(dashed straight line).}\label{1}
\end{figure}

\section{GME for $(d\times d\times d)$-dimensional quantum states}
Next we consider the separability and GME of $(d\times d\times d)$-dimensional quantum states.
Let $\lambda_{i}^{(j)}$ $(j=1,2,3, i=1,\cdot\cdot\cdot,d^{2}-1)$ denote the generators of the special unitary Lie group $SU(d)$
associated with the $j$th $d$-dimensional Hilbert space with orthonormal basis $\{|a\rangle\}_{a=0}^{d-1}$ \cite{dj},\\
for $i=1, \cdots, d-1,$
\begin{center}
$\lambda_{i}^{(j)}=\sqrt{\frac{2}{i(i+1)}}(\sum\limits_{a=0}^{i-1}|a\rangle\langle a|-i|i\rangle\langle i|)$
\end{center}
for $i=d, \cdots, \frac{(d+2)(d-1)}{2},$
\begin{center}
$\lambda_{i}^{(j)}=|j\rangle\langle k|+|k\rangle\langle j|$
\end{center}
when $i=\frac{d(d+1)}{2},\cdots,d^{2}-1,$
\begin{center}
$\lambda_{i}^{(j)}=-i(|j\rangle\langle k|-|k\rangle\langle j|),$
\end{center}
where $0\leq j<k\leq d-1$.

A general tripartite quantum state $\rho\in H_1^d\otimes H_2^d\otimes H_3^d$ can be written in the following Bloch representation,
\begin{equation}\label{14}
\begin{split}
\rho=&\frac{1}{d^{3}}I_{d}\otimes I_{d}\otimes I_{d}+\frac{1}{2d^{2}}(\sum\limits_{i=1}^{d^{2}-1}t_{i}^{1}\lambda_{i}^{(1)}\otimes I_{d}\otimes I_{d}+\sum\limits_{j=1}^{d^{2}-1}t_{j}^{2}I_{d}\otimes \lambda_{j}^{(2)}\otimes I_{d}+\sum\limits_{k=1}^{d^{2}-1}t_{k}^{3}I_{d}\\
&\otimes I_{d}\otimes\lambda_{k}^{(3)})+\frac{1}{4d}(\sum\limits_{i,j=1}^{d^{2}-1}t_{ij}^{12}\lambda_{i}^{(1)}\otimes\lambda_{j}^{(2)}\otimes I_{d}+\sum\limits_{i,k=1}^{d^{2}-1}t_{ik}^{13}\lambda_{i}^{(1)}\otimes I_{d}\otimes\lambda_{k}^{(3)}\\
&+\sum\limits_{j,k=1}^{d^{2}-1}t_{jk}^{23}I_{d}\otimes\lambda_{j}^{(2)} \otimes\lambda_{k}^{(3)})+\frac{1}{8}\sum\limits_{i,j,k=1}^{d^{2}-1}t_{ijk}\lambda_{i}^{(1)}\otimes\lambda_{j}^{(2)}\otimes\lambda_{k}^{(3)},
\end{split}
\end{equation}\\
where $I_{d}$ denotes the $d\times d$ identity matrix,
$t_{i}^{1}=\mbox Tr(\rho\lambda_{i}^{(1)}\otimes I_d\otimes I_d)$, $t_{j}^{2}=\mbox Tr(\rho I_d\otimes \lambda_{j}^{(2)}\otimes I_d)$, $t_{k}^{3}=\mbox Tr(\rho I_d\otimes I_d\otimes \lambda_{k}^{(3)})$, $t_{ij}^{12}=\mbox Tr(\rho\lambda_{i}^{(1)}\otimes\lambda_{j}^{(2)}\otimes I_d)$, $t_{ik}^{13}=\mbox Tr(\rho\lambda_{i}^{(1)}\otimes I_d\otimes \lambda_{k}^{(3)})$, $t_{jk}^{23}=\mbox Tr(\rho I_d\otimes\lambda_{j}^{(2)}\otimes\lambda_{k}^{(3)})$ and $t_{ijk}=\mbox Tr(\rho\lambda_{i}^{(1)}\otimes\lambda_{j}^{(2)}\otimes\lambda_{k}^{(3)})$.
Let $T^{1|23}_{1}$, $T^{1|23}_{2}$, $T^{1|23}_{3}$, $T^{2|13}$, $T^{3|12}_{1}$, $T^{3|12}_{2}$ and $T^{3|12}_{3}$ denote the matrices with entries $t_{1jk}$, $t_{djk}$, $t_{\frac{d(d+1)}{2}jk}$, $t_{i1k}$, $t_{ij1}$, $t_{ijd}$ and $t_{ij\frac{d(d+1)}{2}}$$(i,j,k=1,2,3)$, respectively.
Set $N^{1|23}=15T_{1}^{1|23}+T_{2}^{1|23}+T_{3}^{1|23}$, $N^{2|13}=4T^{2|13}$, $N^{3|12}=15T_{1}^{3|12}+T_{2}^{3|12}+T_{3}^{3|12}$ and $T(\rho)=\frac{1}{3}(\|N^{1|23}\|_{tr}+\|N^{2|13}\|_{tr}+\|N^{3|12}\|_{tr})$.

\begin{lemma}\label{lemma:2} If the state $\rho\in H_{1}^{d}\otimes H_{2}^{d}\otimes H_{3}^{d}$ is a bipartite separable pure state, corresponding to the case (i) and (ii), we have\\
(1) If $\rho$ is separable under bipartition $1|23$, then $\|N^{1|23}\|_{tr}\leq\sqrt{(2-\frac{2}{d})[15^2(2-\frac{2}{d})+2]}$~~or\\
$\sqrt{233(2-\frac{2}{d})}+15\sqrt{(1-\frac{2}{d})(2-\frac{2}{d})}$;\\
(2) If $\rho$ is separable under bipartition $2|13$, then $\|N^{2|13}\|_{tr}\leq4[4\sum\limits_{k=1}^{d-2}\sqrt{\frac{1}{k(k+1)}}
\sqrt{\frac{1}{k+1}-\frac{1}{d}}+\frac{2(d^2-1)}{d}]$\\
or~~$4\sqrt{2-\frac{2}{d}}(1+\sqrt{1-\frac{2}{d}})$;\\
(3) If $\rho$ is separable under bipartition $3|12$, then $\|N^{3|12}\|_{tr}\leq\sqrt{(2-\frac{2}{d})[15^2(2-\frac{2}{d})+2]}$~~or\\
$\sqrt{233(2-\frac{2}{d})}+15\sqrt{(1-\frac{2}{d})(2-\frac{2}{d})}$;
\end{lemma}

{\it Proof}~~(1) If $\rho=|\varphi\rangle\langle\varphi|$ is separable under the bipartition $1|23$, i.e., $|\varphi\rangle=|\varphi_{1}\rangle\otimes|\varphi_{23}\rangle\in H_{1}^{d}\otimes H_{23}^{d^2}$, by Schmidt decomposition we have
\begin{equation}\label{15}
 |\varphi\rangle=\tau_{0}|0a_0\rangle+\tau_{1}|1a_1\rangle+\cdots+\tau_{d-1}|d-1,a_{d-1}\rangle
\end{equation}
with $\sum\limits_i|\tau_{i}|^{2}=1$.
Taking into account the local unitary equivalence in $H_2^d\otimes H_3^d$, if $|\varphi_{23}\rangle\in H_{23}^{d^2}$ is separable we can transform $\{|a_0\rangle,  |a_1\rangle, \cdots, |a_{d-1}\rangle\}$ into the orthonormal bases which constitute separable states: (i) $\{|a_0\rangle, |a_1\rangle, \cdots, |a_{d-1}\rangle\}=\{|00\rangle, |01\rangle, \cdots, |0,d-1\rangle\}$, i.e. $|\varphi_{1|23}\rangle=|\varphi_{1}\rangle\otimes|\varphi_{2}\rangle\otimes|\varphi_{3}\rangle$ is fully separable. When $|\varphi_{23}\rangle\in H_{23}^{d^2}$ is entangled, we can transform $\{|a_0\rangle, |a_1\rangle, \cdots, |a_{d-1}\rangle\}$ into the orthonormal bases which constitute entangled states: (ii) $\{|a_0\rangle, |a_1\rangle, \cdots, |a_{d-1}\rangle\}=\{|00\rangle, |11\rangle, \cdots, |d-1,d-1\rangle\}$.

In the first case, $|\varphi\rangle=\tau_{0}|000\rangle+\tau_{1}|101\rangle+\cdots+\tau_{d-1}|d-1,0,d-1\rangle$, the matrices $T_{1}^{1|23}$, $T_{2}^{1|23}$ and $T_{3}^{1|23}$ are given by\\
$$T_1^{1|23}=\left[\small
               \begin{array}{cccccc}
                 \tau_{0}^2+\tau_{1}^2 & \sqrt{\frac{1}{3}}(\tau_{0}^2+\tau_{1}^2) & \cdots & \sqrt{\frac{2}{(d-1)d}}(\tau_{0}^2+\tau_{1}^2) & \cdots & 0 \\
                 \sqrt{\frac{1}{3}}(\tau_{0}^2-\tau_{1}^2) & \frac{1}{3}(\tau_{0}^2-\tau_{1}^2) & \cdots & \sqrt{\frac{2}{3(d-1)d}}(\tau_{0}^2-\tau_{1}^2) & \cdots & 0 \\
                 \vdots & \vdots & \  & \vdots & \  & \vdots \\
                 \sqrt{\frac{2}{(d-1)d}}(\tau_{0}^2-\tau_{1}^2) & \sqrt{\frac{2}{3(d-1)d}}(\tau_{0}^2-\tau_{1}^2) & \cdots & \frac{2}{(d-1)d}(\tau_{0}^2-\tau_{1}^2) & \cdots & 0 \\
                 \vdots & \vdots & \  & \vdots & \  & \vdots \\
                 0 & 0 & \cdots & 0 & \cdots & 0 \\
               \end{array}
             \right],$$
$$T_{2}^{1|23}=
\left[
                 \begin{array}{cccccc}
                   0 & 0 & \cdots & 0 & \cdots & 0 \\
                   \vdots & \vdots & \  & \vdots & \  & \vdots \\
                   0 & 0 & \cdots & 0 & \cdots & 0 \\
                   2\tau_{0}\tau_{1} & \frac{2\tau_{0}\tau_{1}}{\sqrt{3}} & \cdots & \frac{2\sqrt{2}\tau_{0}\tau_{1}}{\sqrt{(d-1)d}} & \cdots & 0 \\
                   \vdots & \vdots & \  & \vdots & \  & \vdots \\
                   0 & 0 & \cdots & 0 & \cdots & 0 \\
                   \vdots & \vdots  & \  & \vdots & \  & \vdots \\
                   0 & 0 & \cdots & 0 & \cdots & 0 \\
                 \end{array}
               \right],
T_{3}^{1|23}=\left[
                 \begin{array}{cccccc}
                   0 & 0 & \cdots & 0 & \cdots & 0 \\
                   \vdots & \vdots & \  & \vdots & \  & \vdots \\
                   0 & 0 & \cdots & 0 & \cdots & 0 \\
                   0 & 0 & \cdots & 0 & \cdots & 0 \\
                   \vdots & \vdots & \  & \vdots & \  & \vdots \\
                   -2\tau_{0}\tau_{1} &\frac{-2\tau_{0}\tau_{1}}{\sqrt{3}} & \cdots & \frac{-2\sqrt{2}\tau_{0}\tau_{1}}{\sqrt{(d-1)d}} & \cdots & 0 \\
                   \vdots & \vdots  & \  & \vdots & \  & \vdots \\
                   0 & 0 & \cdots & 0 & \cdots & 0 \\
                 \end{array}
               \right].$$\\
Thus
\begin{equation}\label{17}
\begin{split}
\|N^{1|23}\|_{tr}=&\||\alpha\rangle\langle\beta|\|_{tr}\\
=&\sqrt{15^2[(\tau_{0}^2+\tau_{1}^2)^2+(1-\frac{2}{d})(\tau_{0}^2-\tau_{1}^2)^2]+8\tau_{0}^2\tau_{1}^2}\sqrt{2-\frac{2}{d}}\\
\leq&\sqrt{(2-\frac{2}{d})[15^2(2-\frac{2}{d})+2]},
\end{split}
\end{equation}
where
$$|\alpha\rangle=\left[
                   \begin{array}{ccccccccc}
                     15(\tau_{0}^2+\tau_{1}^2) & 15\sqrt{\frac{1}{3}}(\tau_{0}^2-\tau_{1}^2) & \cdots & 15\sqrt{\frac{2}{(d-1)d}}(\tau_{0}^2-\tau_{1}^2) & 2\tau_0^2\tau_1^2 & \cdots & -2\tau_0^2\tau_1^2 & \cdots & 0 \\
                   \end{array}
                 \right]^t,$$
$$|\beta\rangle=\left[
                  \begin{array}{ccccccc}
                    1 & \sqrt{\frac{1}{3}} & \cdots & \sqrt{\frac{2}{(d-1)d}} & 0 & \cdots & 0\\
                  \end{array}
               \right]^t,$$
we have used $\||\alpha\rangle\langle \beta|\|_{tr}=\||\alpha\rangle\|\||\beta\rangle\|$ for vectors $|\alpha\rangle$ and $|\beta\rangle$.

In the second case,   $|\varphi\rangle=\tau_{0}|000\rangle+\tau_{1}|111\rangle+\cdots+\tau_{d-1}|d-1,d-1,d-1\rangle$, the matrices $T_{1}^{1|23}$, $T_{2}^{1|23}$ and $T_{3}^{1|23}$ are given by\\
$$T_1^{1|23}=\left[
               \begin{array}{cccccc}
                 \tau_{0}^2-\tau_{1}^2 & \sqrt{\frac{1}{3}}(\tau_{0}^2+\tau_{1}^2) & \cdots & \sqrt{\frac{2}{(d-1)d}}(\tau_{0}^2+\tau_{1}^2) & \cdots & 0 \\
                 \sqrt{\frac{1}{3}}(\tau_{0}^2+\tau_{1}^2) & \frac{1}{3}(\tau_{0}^2-\tau_{1}^2) & \cdots & \sqrt{\frac{2}{3(d-1)d}}(\tau_{0}^2-\tau_{1}^2) & \cdots & 0 \\
                 \vdots & \vdots & \  & \vdots & \  & \vdots \\
                 \sqrt{\frac{2}{(d-1)d}}(\tau_{0}^2+\tau_{1}^2) & \sqrt{\frac{2}{3(d-1)d}}(\tau_{0}^2-\tau_{1}^2) & \cdots & \frac{2}{(d-1)d}(\tau_{0}^2-\tau_{1}^2) & \cdots & 0 \\
                 \vdots & \vdots & \  & \vdots & \  & \vdots \\
                 0 & 0 & \cdots & 0 & \cdots & 0 \\
               \end{array}
             \right],$$
$$T_{2}^{1|23}=\left[
                  \begin{array}{ccccccc}
                    0 & \cdots & 0 & \cdots & 0 & \cdots & 0 \\
                    \vdots & \  & \vdots & \  & \vdots & \  & \vdots \\
                    0 & \cdots & 2\tau_{0}\tau_{1} & \cdots & 0 & \cdots & 0 \\
                    \vdots & \  & \vdots & \  & \vdots & \  & \vdots \\
                    0 & \cdots & 0 & \cdots & -2\tau_{0}\tau_{1} & \cdots & 0 \\
                    \vdots & \  & \vdots & \  & \vdots & \  & \vdots \\
                    0 & \cdots & 0 & \cdots & 0 & \cdots & 0 \\
                  \end{array}
                \right],
T_{3}^{1|23}=\left[
                  \begin{array}{ccccccc}
                    0 & \cdots & 0 & \cdots & 0 & \cdots & 0 \\
                    \vdots & \  & \vdots & \  & \vdots & \  & \vdots \\
                    0 & \cdots & 0 & \cdots & -2\tau_{0}\tau_{1} & \cdots & 0 \\
                    \vdots & \  & \vdots & \  & \vdots & \  & \vdots \\
                    0 & \cdots & -2\tau_{0}\tau_{1} & \cdots & 0 & \cdots & 0 \\
                    \vdots & \  & \vdots & \  & \vdots & \  & \vdots \\
                    0 & \cdots & 0 & \cdots & 0 & \cdots & 0 \\
                  \end{array}
                \right].$$
Then
\begin{equation}\label{18}
N^{1|23}=15(F+|\gamma\rangle\langle\zeta|)+T_2^{1|23}+T_3^{1|23},
\end{equation}
where $F$ is a $(d^2-1)$-dimensional square matrix which the first column is same as the first column of $T_1^{1|23}$ and the other elements are zero,
$|\gamma\rangle=\footnotesize\left[
                   \begin{array}{cccccc}
                     (\tau_{0}^2+\tau_{1}^2) & \sqrt{\frac{1}{3}}(\tau_{0}^2-\tau_{1}^2) & \cdots & \sqrt{\frac{2}{(d-1)d}}(\tau_{0}^2-\tau_{1}^2) & \cdots & 0 \\
                   \end{array}
                 \right]^t$,
$|\zeta\rangle=\footnotesize\left[
                  \begin{array}{ccccccc}
                    0 & \sqrt{\frac{1}{3}} & \cdots & \sqrt{\frac{2}{(d-1)d}} & 0 & \cdots & 0 \\
                  \end{array}
               \right]^t$.\\
Thus
\begin{equation}\label{19}
\begin{split}
\|N^{1|23}\|_{tr}\leq&15(\|F\|_{tr}+\||\gamma\rangle\|\||\zeta\rangle\|)+\|T_2^{1|23}+T_3^{1|23}\|_{tr}\\
\leq&15\sqrt{(1-2\tau_{1}^2)^2+(1-\frac{2}{d})}+15\sqrt{(1-\frac{2}{d})(2-\frac{2}{d})}+4\sqrt{2}\sqrt{\tau_{1}^2(1-\tau_{1}^2)}\\
\leq&\sqrt{233(2-\frac{2}{d})}+15\sqrt{(1-\frac{2}{d})(2-\frac{2}{d})}.
\end{split}
\end{equation}
where we have used $\|A+B\|_{tr}\leq\|A\|_{tr}+\|B\|_{tr}$ for matrices $A$ and $B$ and the upper bound is obtained by taking the extreme value of the function with independent variable $\tau_{1}$.\\
(2) If $\rho=|\varphi\rangle\langle\varphi|$ is separable under the bipartition $2|13$, i.e., $|\varphi\rangle=|\varphi_{2}\rangle\otimes|\varphi_{13}\rangle\in H_{2}^{d}\otimes H_{13}^{d^2}$. We only need to consider two cases:\\
(i) $|\varphi\rangle=\tau_{0}|000\rangle+\tau_{1}|101\rangle+\cdots+\tau_{d-1}|d-1,0,d-1\rangle$.
$$T^{2|13}=\footnotesize\left[
                   \begin{array}{ccccccccc}
                     \tau_{0}^2+\tau_{1}^2 & \sqrt{\frac{1}{3}}(\tau_{0}^2-\tau_{1}^2) & \cdots & \sqrt{\frac{2}{(d-1)d}}(\tau_{0}^2-\tau_{1}^2) & \  & \  & \  &  \ & \  \\
                     \sqrt{\frac{1}{3}}(\tau_{0}^2-\tau_{1}^2) & \frac{1}{3}(\tau_{0}^2+\tau_{1}^2+4\tau_2^2) & \cdots & \sqrt{\frac{2}{3(d-1)d}}(\tau_{0}^2+\tau_{1}^2-2\tau_2^2) & \  & \  & \  & \  & \  \\
                     \vdots & \vdots & \  & \vdots & \  & \  & \  & \  & \  \\
                     \sqrt{\frac{2}{(d-1)d}}(\tau_{0}^2-\tau_{1}^2) & \sqrt{\frac{2}{3(d-1)d}}(\tau_{0}^2+\tau_{1}^2-2\tau_2^2) & \cdots & \frac{2}{(d-1)d}(\sum\limits_{i=0}^{d-2}\tau_{i}^2+(d-1)^2\tau_{d-1}^2) & \  & \  & \  & \  & \ \\
                     \  & \  & \  & \  &\ddots& \  & \  & \  & \  \\
                     \  & \  & \  & \  & 2\tau_k\tau_j& \  & \  & \  \\
                     \  & \  & \  & \  & \  & \ddots& \  & \  \\
                     \  & \  & \  & \  & \  & \  & -2\tau_k\tau_j& \  \\
                     \  & \  & \  & \  & \  & \  & \  & \ddots\\
                   \end{array}
                 \right],$$
where $0\leq k<j\leq d-1$.
Then
\begin{equation}\label{20}
\begin{split}
\|N^{2|13}\|_{tr}\leq&4[4\sum\limits_{k=1}^{d-2}\sqrt{\frac{1}{k(k+1)}}
\sqrt{\frac{1}{k+1}-\frac{1}{d}}|\sum\limits_{i=0}^{k-1}\tau_{i}^2-k\tau_{k}^2|
+4\sum\limits_{0\leq i<j\leq d-1}|\tau_i\tau_j|\\
&+\sum\limits_{k=1}^{d-1}\frac{2}{k(k+1)}|\sum\limits_{i=0}^{k-1}\tau_{i}^2+k^2\tau_{k}^2|]\\
\leq&4[4\sum\limits_{k=1}^{d-2}\sqrt{\frac{1}{k(k+1)}}
\sqrt{\frac{1}{k+1}-\frac{1}{d}}+4\frac{(d-1)\sum_{i=0}^{d-1}\tau_{i}^2}{2}+2(1-\frac{1}{d})\sum\limits_{i=0}^{d-1}\tau_{i}^2]\\
=&4[4\sum\limits_{k=1}^{d-2}\sqrt{\frac{1}{k(k+1)}}
\sqrt{\frac{1}{k+1}-\frac{1}{d}}+\frac{2(d^2-1)}{d}],
\end{split}
\end{equation}
where we have used $\|A+B\|_{tr}\leq\|A\|_{tr}+\|B\|_{tr}$.

(ii) $|\varphi\rangle=\tau_{0}|000\rangle+\tau_{1}|111\rangle+\cdots+\tau_{d-1}|d-1,d-1,d-1\rangle$.

We have $T^{2|13}=T_1^{1|23}=F+|\gamma\rangle\langle\zeta|$, then
\begin{equation}\label{21}
\begin{split}
\|N^{2|13}\|_{tr}=&4[\sqrt{(\tau_{0}^2-\tau_{1}^2)^2+(1-\frac{2}{d})(\tau_{0}^2+\tau_{1}^2)^2}+
\sqrt{(1-\frac{2}{d})((\tau_{0}^2+\tau_{1}^2)^2+(1-\frac{2}{d})(\tau_{0}^2-\tau_{1}^2)^2)}]\\
\leq&4\sqrt{2-\frac{2}{d}}(1+\sqrt{1-\frac{2}{d}}).
\end{split}
\end{equation}
(3) If $\rho=|\varphi\rangle\langle\varphi|$ is separable under the bipartition $3|12$, i.e., $|\varphi\rangle=|\varphi_{3}\rangle\otimes|\varphi_{12}\rangle\in H_{3}^{d}\otimes H_{12}^{d^2}$.

In the first case, we have $T_1^{3|12}=(T_1^{1|23})^t$, $T_2^{3|12}=(T_2^{1|23})^t$, $T_3^{3|12}=(T_3^{1|23})^t$, therefore
$$\|N^{3|12}\|_{tr}=\|(N^{1|23})^t\|_{tr}\leq\sqrt{(2-\frac{2}{d})[15^2(2-\frac{2}{d})+2]}.$$

In the second case, we have $T_1^{3|12}=T_1^{1|23}$, $T_2^{3|12}=T_2^{1|23}$, $T_3^{3|12}=T_3^{1|23}$, therefore
$$\|N^{3|12}\|_{tr}=\|N^{1|23}\|_{tr}\leq\sqrt{233(2-\frac{2}{d})}+15\sqrt{(1-\frac{2}{d})(2-\frac{2}{d})}.$$

{\bf Remark 3.} After the Schmidt
decomposition of a general state, we have $|\varphi_{1|23}\rangle=|\varphi_{1}\rangle\otimes|\varphi_{23}\rangle=\tau_{0}|0a_0\rangle+\cdots+\tau_{d-1}|d-1,a_{d-1}\rangle$, under LU equivalence we choose $\{|a_0\rangle, \cdots, |a_{d-1}\rangle\}=\{|00\rangle, \cdots, |0,d-1\rangle\}$ in the case (i) to transform $\{|a_0\rangle, \cdots, |a_{d-1}\rangle\}$ into the orthonormal basis elements which constitute
separable states, then $|\varphi_{23}\rangle$ must be a separable state such that $|\varphi_{23}\rangle=|\varphi_2\rangle\otimes|\varphi_{3}\rangle$, which implies that $|\varphi\rangle$ is fully separable, then we have $\|N^{1|23}\|_{tr}\leq\sqrt{(2-\frac{2}{d})[15^2(2-\frac{2}{d})+2]}$. Consequently, if $\|N^{1|23}\|_{tr}>\sqrt{(2-\frac{2}{d})[15^2(2-\frac{2}{d})+2]}$, $\rho$ is not fully separable. Therefore, Lemma 2 can also be used to detect not fully separable.

\begin{theorem}\label{thm:2} For a mixed state $\rho\in H_{1}^{d}\otimes H_{2}^{d}\otimes H_{3}^{d}$, if it holds that
\begin{equation}\label{22}
T(\rho)=\frac{1}{3}(\|N^{1|23}\|_{tr}+\|N^{2|13}\|_{tr}+\|N^{3|12}\|_{tr})>M,
\end{equation}
then $\rho$ is genuine multipartite entangled, where $M=\mbox max\{\sqrt{(2-\frac{2}{d})[15^2(2-\frac{2}{d})+2]}, \sqrt{233(2-\frac{2}{d})}+15\sqrt{(1-\frac{2}{d})(2-\frac{2}{d})}, 4[4\sum\limits_{k=1}^{d-2}\sqrt{\frac{1}{k(k+1)}}
\sqrt{\frac{1}{k+1}-\frac{1}{d}}+\frac{2(d^2-1)}{d}]\}$.
\end{theorem}
{\it Proof}~~If a mixed state $\rho=\sum p_{i}\rho_{i}$, $\sum p_{i}=1$, is biseparable, by using Lemma 2, we have
\begin{equation*}
\begin{split}
T(\rho)=&\frac{1}{3}(\|N^{1|23}(\rho)\|_{tr}+\|N^{2|13}(\rho)\|_{tr}+\|N^{3|12}(\rho)\|_{tr})\\
\leq&\frac{1}{3}\sum p_{i}(\|N^{1|23}(\rho_{i})\|_{tr}+\|N^{2|13}(\rho_{i})\|_{tr}+
\|N^{3|12}(\rho_{i})\|_{tr})\\
\leq&\frac{1}{3}(M+M+M)\\
=&M.
\end{split}
\end{equation*}
Consequently, if $T(\rho)>M$, $\rho$ is GME. \qed

{\bf Remark 4.} Using a different method and the matrix norm,
the authors \cite{vh} considered the Ky Fan norm of the matricizations of tensors to derive GME conditions for tripartite qubits and four partite qubits. While in our current approach we have used the correlation tensors in the Bloch representation of density matrices coupled with a linear combination of the special matrices, our main structural matrices in the algorithm differ from other references both in structure and in form. We have not only employed the trace norm but also applied local unitary equivalence and the Schmidt decompositions to study the GME of arbitrary dimensional quantum systems. It turns out that our special matrices of
correlation tensors constructed in this way greatly simplify the criteria operationally, as shown in Example 1 that our theorem is more effective than Theorem 2 in \cite{vh} for $d=2$.

\textit{\textbf{Example 2}} Consider the mixed state in three-qutrit quantum system,
\begin{equation}\label{23}
\rho=\frac{1-x}{27}I_{27}+x|W(3)\rangle\langle
W(3)|,~~~0\leq x\leq1,
\end{equation}
where $|W(3)\rangle=\frac{1}{\sqrt{6}}(|001\rangle+|010\rangle+|100\rangle
+|112\rangle+|121\rangle+|211\rangle)$ is the $3\times 3 \times 3$ $W$ state, $I_{27}$ is $27\times 27$ identity matrix.
For $d=3$, from Theorem 1 in \cite{vh} and our Theorem 2, we have $f_4(x)=2.372684x-2.177324$ and $f_5(x)=T(\rho)-M=34.5797x-27.6257$, respectively. When $f_4(x)>0$, Theorem 1 in \cite{vh} detects the GME for $0.917663<x\leq1$, while our Theorem 2 detects the GME for $0.798899<x\leq1$, see Fig. 2, which shows that our result (Theorem 2) is able to detect more genuine tripartite entangled states.
\begin{figure}
  \centering
  \includegraphics[width=8cm]{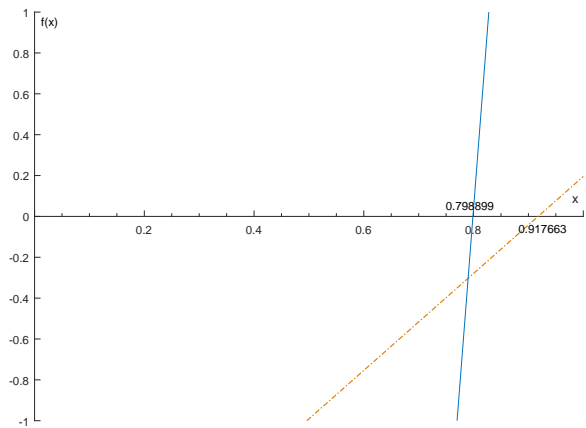}\\
  \caption{Detecting the GME of $\rho$ in Example 2. $f_4(x)$ from the Theorem 1 in \cite{vh} (dash-dot straight line), $f_5(x)$ from our Theorem 2 (solid straight line).}
\end{figure}

\section{ Conclusions}
We have studied the entanglement and genuine multipartite entanglement in tripartite quantum systems.
The general state has been transformed into a simpler form by the Schmidt decomposition and local unitary transformation, then we have constructed some new
operational matrices and considered the trace norm of the sum of these matrices, from which we have obtained new criteria in detecting tripartite entanglement for $(2\times2\times 2)$ and general $(d\times d\times d)$ systems. Detailed examples show that our criteria are able to detect genuine tripartite entanglement
more effective
than some existing criteria. Our approach can be also applied to general multipartite systems.

\noindent\textbf {Acknowledgements}
This work is supported by the National Natural Science Foundation of China under grant nos. 11101017, 11531004, 11726016 and 12075159,
Simons Foundation under grant no. 523868, Beijing Natural Science Foundation (grant no. Z190005), Academy for Multidisciplinary Studies, Capital Normal University, and Shenzhen Institute for Quantum Science and Engineering, Southern University of Science and Technology (no. SIQSE202001).

\noindent\textbf {Data Availability Statements} All data generated or analysed during this study are available from the corresponding author on reasonable request.

\end{document}